\documentstyle{elsart}
\begin{document}
\begin{frontmatter}

\title{The Non-Abelian Topological Gauge \\ Field Theory of $\tilde{p}$--Branes \thanksref{thankslabel}}
\author[lzu]{Yi-Shi Duan},
\author[lzu]{Ji-Rong Ren\corauthref{cor}}   \corauth[cor]{Corresponding author} \ead{renjr@lzu.edu.cn}
\address[lzu]{Institute of Theoretical Physics, \\      School of Physical Science and Technology,\\ Lanzhou University, Lanzhou, 730000,  P. R.  China}
\thanks[thankslabel]{This work was supported by the National Natural Science Foundation of China and the Doctoral Foundation from the Ministry of
Education of China. Dedicated to Prof. Shiing-Shen Chern for his contribution.}

\begin{abstract}
By the generalization of Chern--Simons topological current and Gauss--Bonnet-Chern theorem, the purpose of this paper is to make a non-Abelian gauge field theory foundation of the topological current of $\tilde{p}$-branes formulated in our previous work. Using $\phi $--mapping topological current theory proposed by Professor Duan, we find that the topological $\tilde p$-branes are created at every isolated zero of vector field $\vec \phi (x)$. It is shown that the  topological charges carried by $\tilde p$-branes are topologically quantized and labeled by Hopf index
and Brouwer degree, i.e., the winding number of the $\phi
$--mapping. The action of topological $\tilde p$--branes is
obtained and is just Nambu action for multistrings when $D- \tilde d=2$.
\end{abstract}

\end{frontmatter}


\section{Introduction}
Extended objects with $p$ spatial dimension, known as $
p$-branes, play an essential role in revealing the nonperturbative
structure of the superstring theories and
M--theories\cite{b1,b2,b3,b4,StelleTownsend1987,SchwarzSeibergRMP1999}.
Antisymmetric tensor gauge
fields determine all of the features of a $ p$--brane and
have been widely studied in the theory of $ p$--branes
\cite{DiamantiniPLB1996,a1,a2,a3,duff}. In the context of the effective $D=10$ or $D=11$
supergravity theory a $p$-brane is a $p$-dimensional extended source for a $%
(p+2)$-form gauge field strength $F.$ It is well-known that the $(p+2)$-form
strength $F$ satisfies the field equation%
\begin{equation} \label{No.1}
\nabla _\mu F^{\mu \mu _1\cdots \mu _{p+1}}=j^{\mu _1\cdots \mu _{p+1}}
\end{equation}
where $j^{\mu _1\cdots \mu _{p+1}}$is a $(p+1)$-form tensor current and corresponding to "electric source", and the dual field strength $^{*}F$ satisfies%
\begin{equation}\label{TCCTM}
\nabla _\mu ~^{*}F^{\mu \mu _1\cdots \mu _{\tilde p+1}}=\tilde j^{\mu
_1\cdots \mu _{\tilde p+1}}
\end{equation}
in which $\tilde j^{\mu _1\cdots \mu _{\tilde p+1}}$ is a extended  $(\tilde p+1)$-form topological  tensor current and corresponding to "magnetic source" \cite{strom,duff,hull}. In Ref.\cite{DiamantiniPLB1996,strom,duff,DvaliPRD2000,CembranosPRD2002},
from the perspective of a higher dimensional theory, the topological
theories of $\tilde  p$--branes in $M$--theory were also studied. The $\phi $--mapping topological
current theory proposed by Professor Duan plays a crucial role in
studying the structure of topological defects\cite{DuanSlac1984,DuanRenYang2003,DuanLiuJHEP2004,DuanMengJMP1993,DuanLiYangNPB1998,DuanFuJMP1998,DuanLiuFuPRD2003,DUANZHangLiPRB1998,DUANLiuZHangJP2002,DuanZhangFuPRE1999,DuanZhangPRE1999,DuanFuZhangPRD2000,DuanFuJiaJMP2000,DuanLiuHopkins1987,DuanZhangMPLA2001}. In our previous work\cite{DuanFuJiaJMP2000}, using the $\phi $--mapping topological current theory, we had present a new topological tensor current of $\tilde p$--branes. It's shown that the current is identically conserved and behaves as $\delta (\vec \phi )$, and every isolated zero of the  field $\vec \phi (x)$ corresponding to a "magnetic" $\tilde p$--brane. It must be pointed out that usually the study of extended $\tilde p$--branes is always via generalizing Kalb-Ramond Abelian gauge field\cite{DiamantiniPLB1996,NepomechiePRD1985}. That is a kind of $U(1)$ gauge field theory, so far from which the topological  current  can't be strictly induced. In fact the present work is a generalization of $GBC$ topological current of moving point defect\cite{DuanMengJMP1993,DuanLiYangNPB1998}. By making use of the generalization of Gauss-Bonnet-Chern theorm and $\phi $--mapping field theory, we find a $SO(N)$ non-Abelian topological field theory of $\tilde p$--branes, in which the dual field strength $^* F^{\mu \mu _1 \cdots \mu _{\tilde p +1}}$ of eq.(\ref{TCCTM}) can  rigorously create a topological current of $\tilde p$--branes in a natural way.
In the $SO(N)$ non-Abelian topological gauge field theory of $\tilde p$--branes, we also investigate the inner  structure of topological current of $\tilde p$--branes,  show that the topological charges of $\tilde p$--branes  are topologically quantized and labeled by the Hopf index and Brouwer degree, the winding number of the $\phi $--mapping. We also find that in this $SO(N)$ gauge field theory of $\tilde p$--branes, when $N$ is even, the topological tensor current of $\tilde p$--branes $\tilde j^{\mu _1\cdots \mu _{\tilde p+1}} $ can be looked upon as  the generalization of Chern-Simoin topological current, that we have formulated in \cite{DuanFuJiaJMP2000}.

\section{The non-Abelian field theory of $\tilde{p}$--branes}

In this paper the studies of the non-Abelian gauge field
theory and topological current of $\tilde{p}$--branes are basing
on the Gauss--Bonnet--Chern($GBC$) theorem and the generalization
of Chern--Simons topological current. It is well-known that Gauss-Bonnet-Chern theorem is a generalization of Euler number density from two dimensional
Gauss--Bonnet theorem to arbitrary even dimensional theory,
which relates the curvature of the compact and oriented
even--dimensional Riemannian manifold $M$ with an important
topological invariant, the Euler-Poincar\'e characteristic $\chi
{(M)}$. The $GBC$--form
corresponding to Euler number density is given by
\begin{equation}\label{GBCform1}
\Lambda =  {\frac{(-1)^{\frac{N}{2}-1}}{2^N\pi ^{\frac
N2}\left(\frac{N}{2}\right)!}} \varepsilon_{A_1A_2\cdots
{A_{N-1}A_N}}F^{{A}_1{A}_2}\wedge \cdots \wedge
{F^{{A}_{N-1}{A}_N}},
\end{equation}
in which $F^{AB}$ is the curvature tensor of $SO(N)$ principal bundle of the Riemannian manifold $M$, i.e., the $SO(N)$ gauge field  2--form
\begin{equation}\label{}
F^{AB}=d\omega ^{AB}-\omega ^{AC}\wedge \omega ^{CB}.
\end{equation}
where $\omega ^{AB}$ is the spin connection $1$-form.
In 1944, an elegantly intrinsic proof of the theorem was given by Chern\cite{Cher1}, whose instructive idea was to work on the sphere bundle $S^{N-1}(M)$. Using a recursion method Chern has
proved that the $GBC$--form is exact on $S^{N-1}(M)$
\begin{equation}\label{GBC}
\Lambda =d \Omega ,
\end{equation}
where the $(N-1)$--form $\Omega$ is called the  $Chern$--form \begin{equation}
 \Omega = {1 \over {\left(2\pi \right)^{\frac{N}{2}} }}\sum\limits_{k = 0}^{\frac{N}{2} - 1} {( - 1)^k {{2^{ - k} } \over {(N - 2k - 1)!!k!}}\Theta _k },
\end{equation}
in which
\begin{equation}\label{theta1}
\begin{array}{ll}
\Theta _k =& \varepsilon _{A_1 A_2 \cdots A_{N - 2k} A_{N - 2k + 1} A_{N - 2k + 2} \cdots A_{n - 1} A_N } n^{A_1 } \theta ^{A_2 }\wedge \cdots \vspace{.4cm} \\ &  \wedge \theta ^{A_{N - 2k} } \wedge F^{A_{N - 2k + 1} A_{N - 2k + 2} } \wedge \cdots \wedge F^{A_{N - 1} A_N },
\end{array}
\end{equation}
\begin{equation}
\theta ^A \equiv Dn^A = dn^A - \omega ^{AB} n^B,
\end{equation}
and $n^A$ is the section of the  sphere bundle $S^{N-1}(M)$
\begin{equation}
n:\partial M \to S^{N - 1}(M).
\end{equation}
A detailed review of Chern's proof of the $GBC$ theorem was presented in Ref.\cite{Dowk} and  one great advance in this field is the discovery of the relationship between supersymmetry and the index theorem\cite{Alva}.

It must be pointed out that the $GBC$ theorem is formulated by the exterior differential forms\cite{Cher1}. Differential forms are a vector space (with a C-infinity topology) and therefore have a dual space in higher-dimension space. Submanifolds represent an element of the dual via integration, so it is common to say that they are in the dual space of forms, which is the space of currents\cite{Weissteink-Form}. let $(X,g)$ be a  $D$--dimensional manifold  and $F^{AB}_{\mu \nu }$ the  curvature tensor of $SO(N)$ principal bundle, we can define a $(\tilde p+1)$--dimensional topological tensor current on manifold $X$
\begin{eqnarray}\label{Chernformcompdual1}
\tilde j^{\lambda  \lambda _1 \cdots \lambda _{\tilde p}
}=&&\frac{\varepsilon ^{\lambda \lambda
_1 \cdots \lambda _{\tilde p} \mu _1 \mu _{2} \cdots \mu _{N}}}{\sqrt{g}} {\frac{(-1)^{\frac{N}{2}-1}N!}{2^N (2\pi )^{\frac N
2}\left(\frac{N}{2}\right)!}}
\nonumber \\ && \cdot \varepsilon_{A_1A_2\cdots {A_{N-1}A_N}}
F^{{A}_1{A}_2}_{\mu _1\mu _2 } \cdots F^{{A}_{N-1}{A}_N}_{\mu
_{N-1}\mu _{N}},
\end{eqnarray}
where $N$ is the dimension of a submanifold $M$. It is easy to see that eq.(\ref{Chernformcompdual1}) is just  the generalization of Chern--Simons $SO(2)$ topological current\cite{DuanFuZhangPRD2000}
\begin{equation}\label{}
\tilde j^{\lambda }=\frac{1}{8\pi}
\frac{\varepsilon ^{\lambda \mu \nu}}{\sqrt{g}} \varepsilon_{AB} F^{AB }_{\mu \nu}.
\end{equation}
Using the Bianchi identity,
\begin{equation}\label{Bianchip1}
D _\mu  F_{\nu \lambda }^{AB}  + D _\nu  F_{\lambda \mu }^{AB}  + D _\lambda  F_{\mu \nu }^{AB}  = 0
\end{equation}
one find
\begin{equation}\label{Bianchip2}
\varepsilon ^{\mu _1  \cdots \mu _{i - 1} \mu _i \mu _{i + 1}  \cdots \mu _{N+k} } D _{\mu _{i - 1} } F_{\mu _i \mu _{i + 1} }^{AB}  = 0 .
\end{equation}
From (\ref{Chernformcompdual1}), it can be proved\cite{spivak1975} that
\begin{equation}
\nabla _{\lambda }\tilde j^{ \lambda \lambda _1 \cdots \lambda _{\tilde p}} =0,
\end{equation}
i.e., the antisymmetric topological tensor current   $\tilde j^{ \lambda \lambda _1 \cdots \lambda _{\tilde p}} $ is identically conserved.

It's also easy to find that the topological tensor current $\tilde j^{
\lambda \lambda _1 \cdots \lambda _{\tilde p}} $ is the dual tensor of
$GBC$ tensor defined in eq.(\ref{GBCform1})
\begin{equation}\label{Chernformcompdual2}
\tilde j^{\lambda  \lambda _1 \cdots \lambda _{\tilde p} }=
\frac{\varepsilon ^{\lambda \lambda _1 \cdots \lambda _{\tilde p} \mu _1
\mu _{2} \cdots \mu _{N}}}{\sqrt{g}}\Lambda _{\mu _{1} \mu _{2}
\cdots \mu _{N} }.
\end{equation}
Furthermore, by virtue of the tensor form of $GBC$ theorem (\ref{GBCform1}) and (\ref{GBC})
\begin{equation}\label{CPartofK}
\Lambda _{\mu _{1} \mu _{2} \cdots \mu _{N} } =\partial_{[\mu
_1}F_{\mu _{2} \cdots \mu _{N}] },
\end{equation}
we can  find that the  dual tensor of  $Chern$--tensor
$F_{\mu _{2} \cdots \mu _{N} }$ is
\begin{equation} \label{DualfieldGBC}
{\; ^{\star } F}^{\lambda  \lambda _1 \cdots \lambda _{\tilde p} \mu _1}=
\frac{\varepsilon ^{\lambda \lambda _1 \cdots \lambda _{\tilde p} \mu _1
\mu _{2} \cdots \mu _{N}}}{\sqrt{g}}F_{\mu _{2} \cdots \mu _{N} }.
\end{equation}
It's obvious that if in eq.(\ref{TCCTM}) the dual field tensor   ${\; ^{\star } F}^{\mu  \mu _1 \cdots \mu _{\tilde p +1} }$ is taking the form of (\ref{DualfieldGBC}) deduced from $GBC$ theorem, then we have the conserved topological tensor current (\ref{Chernformcompdual1}).
Therefore for the case of even $N$ the anti-symmetric tensor current(\ref{Chernformcompdual1}) that constructed in term of $SO(N)$ gauge field tensor $F^{AB}_{\mu \nu }$ is just the topological current of creating $\tilde{p}$-branes. As to the field tensor $F^{\mu \mu _1\cdots \mu _{p+1}}$ in (\ref{No.1}) and the dual tensor $~^{*}F^{\mu \mu _1\cdots \mu _{\tilde p+1}} $ in (\ref{TCCTM}) can be found by making use of the Chern-form. In the following we will show that the tensor current defined by (\ref{Chernformcompdual1}) is just the $\phi$--mapping topological tensor current of $\tilde p$--branes in \cite{DuanFuJiaJMP2000}. This is a novel foundation of the non-Abelian topological gauge field theory of $\tilde p$--branes.

The early work of Chern\cite{Cher1} had shown that in a neighborhood of arbitrary
point $P$ on $M$ it can be chosen a family of frames such that the
spin connection $\omega ^{AB}=0$. This  locally Euclidean
homeomorphism immediately gives an important
consequence of the $GBC$ theorem on sphere bundle $S^{N-1}(X)$:
\begin{equation}
\Omega = {1 \over {\left(2\pi \right)^{\frac{N}{2}} }} { {{1 }
\over {(N - 1)!!}}\varepsilon _{A_1 A_2 \cdots A_N } n^{A_1 } dn ^{A_2 }
\wedge \cdots  \wedge dn ^{A_{N} } }. \label{Chernfform2}
\end{equation}
Using the unit sphere area formula $ A(S^{N-1})={{2\pi ^{N/2}}/{{\Gamma (\frac N2)}}} \label{area}
$ and the following relation
\begin{equation}
\left(2\pi \right) ^\frac{N}{2} (N-1)!! = A(S^{N-1})(N-1)!,
\end{equation}
the $Chern$--form expressed by (\ref{Chernfform2})  can be
locally reduced to
\begin{equation}
\Omega = \frac{1}{ A(S^{N-1})(N-1)!} \varepsilon _{A_1 A_2 \cdots
A_N } n^{A_1 } dn ^{A_2 } \wedge \cdots \wedge dn ^{A_{N} }.
\label{Chernform4}
\end{equation}
We see that the expression (\ref{Chernform4}) is nothing but the ratio of area element to the total area $A(S^{N-1})$ of unit sphere $S^{N-1}$. This is essential of $GBC$ theorem.

Using the  $Chern$--form(\ref{Chernform4}), it's easy to prove that the $Chern$ tensor field can be simply written as
\begin{equation}\label{CherenformKcomp}
F_{\mu _{2} \cdots \mu _{N} }= \frac{1}{ A(S^{N-1})(N-1)!}
\varepsilon _{A_1 A_2 \cdots A_N } n^{A_1 } \partial _{\mu _2 }n
^{A_2 }  \cdots \partial _{\mu _N}n ^{A_{N} }
\end{equation}
and the  $(\tilde p+1)$--dimensional tensor current (\ref{Chernformcompdual1}) can also be expressed as follows
\begin{eqnarray} \label{Chernformcompdual3}
\tilde j^{\lambda \lambda _1 \cdots \lambda _{\tilde p}} = &&
\frac{1}{A(S^{N-1})(N-1)!} \varepsilon_{A_1A_2\cdots {A_{N-1}A_N}}
\nonumber \\ && \cdot  \frac{\varepsilon ^{ \lambda  \lambda _1
\cdots \lambda _{\tilde p}  \mu _1 \mu _2 \cdots \mu _N}}{\sqrt{g}}
\partial _{\mu _1}n^{A_1}\cdots  \partial _{\mu _N}n^{A_N}.
\end{eqnarray}
This is just the topological tensor current of $\tilde p$--branes in \cite{DuanFuJiaJMP2000}, i.e., the tensor current of creating $\tilde p$--dimensional manifold\cite{JiangDuanJMP2000,YangJiangDuanCPL2001}.

In case of  $SO(N+1)$ gauge field theory on $D$--dimensional manifold $X$, we can define a new field theory  as
\begin{eqnarray}\label{LambdaPrimeComp1}
F_{\mu _1 \cdots \mu _N} = && {\frac{(-1)^{\frac{N}{2}-1}N!}{2^N
(2\pi )^{\frac N 2}\left(\frac{N}{2}\right)!}}
\varepsilon_{AA_1A_2\cdots {A_{N-1}A_N}} \nonumber \\ && \cdot
n^AF^{{A}_1{A}_2}_{\mu _1\mu _2 } \cdots F^{{A}_{N-1}{A}_N}_{\mu
_{N-1}\mu _{N}},
\end{eqnarray}
\begin{equation}\label{ChernformcompdualN+1}
{{\; ^{\star } F}}^{\lambda  \lambda _1 \cdots \lambda _{\tilde p} \mu }= \frac{\varepsilon ^{\lambda \lambda _1 \cdots \lambda _{\tilde p}  \mu  \mu _1 \mu _{2} \cdots \mu _{N}}}{\sqrt{g}}{F}_{\mu _{1} \mu _{2} \cdots \mu _{N} },
\end{equation}
where $n^A $ is the section of sphere bundle $S^{N}(X)$.
Using (\ref{LambdaPrimeComp1}) and the Bianchi identity (\ref{Bianchip2}), we find that the topological tensor current can be defined as
\begin{equation}\label{TopoTenCurrSON+1}
\begin{array}{ll}
& \displaystyle {\tilde j}^{\lambda  \lambda _1 \cdots \lambda
_{\tilde p}}\\ \vspace{4mm} = & \displaystyle \frac{1}{\sqrt{g}}\partial
_{\mu } \left({\sqrt{g} {\; ^{\star } F}}^{\lambda  \lambda _1
\cdots \lambda _{\tilde p} \mu } \right)  \\\vspace{4mm} = & \displaystyle
\frac{1}{\sqrt{g}}\partial _{\mu } \left(\varepsilon ^{\lambda
\lambda _1 \cdots \lambda _{\tilde p}  \mu  \mu _1 \mu _{2}\cdots \mu
_{N}}F_{\mu _1 \mu _{2}\cdots \mu _{N} } \right)  \\ \vspace{4mm}
= & \displaystyle \frac{\varepsilon ^{\lambda \lambda _1 \cdots
\lambda _{\tilde p}  \mu  \mu _1 \mu _{2}\cdots \mu _{N}}}{\sqrt{g}}D _{\mu
} \left(F_{\mu _1 \mu _{2}\cdots \mu _{N} } \right)  \\
\vspace{4mm} = &\displaystyle {\frac{(-1)^{\frac{N}{2}-1}N!}{2^N
(2\pi )^{\frac N 2}\left(\frac{N}{2}\right)!}}
\varepsilon_{AA_1A_2\cdots {A_{N-1}A_N}} \frac{\varepsilon
^{\lambda  \lambda _1 \cdots \lambda _{\tilde p} \mu \mu _1 \cdots \mu
_{N}}}{\sqrt{g}}\\ &  \cdot D_\mu n^AF^{{A}_1{A}_2}_{\mu _1\mu _2
} \cdots F^{{A}_{N-1}{A}_N}_{\mu _{N-1}\mu _{N}}.
\end{array}
\end{equation}
In the like manner, it can be  proved that the covariant divergence of this  topological tensor current is
\begin{equation}\label{TopoTenCurrSON+1CON}
\begin{array}{ll}
\nabla _{\lambda }  {\tilde j}^{\lambda  \lambda _1 \cdots \lambda
_{\tilde p}}  = &\displaystyle  \frac{\varepsilon ^{\lambda  \lambda _1
\cdots \lambda _{\tilde p} \mu \mu _1 \cdots \mu _{N}}}{\sqrt{g}} \\ &
\cdot  D_{\lambda } D_\mu n^AF^{{A}_1{A}_2}_{\mu _1\mu _2 } \cdots
F^{{A}_{N-1}{A}_N}_{\mu _{N-1}\mu _{N}}.
\end{array}
\end{equation}
Using the definition of  curvature tensor $F^{AB}_{\mu \nu}$ of $SO(N+1)$ principal bundle
\begin{equation}
(D_\mu D_\nu -D_\nu D_\mu)n^A=-F^{AB}_{\mu \nu}n^B,
\end{equation}
we can obtain
\begin{equation}
\begin{array}{ll}
& \nabla _{\lambda } \displaystyle {\tilde j}^{\lambda  \lambda _1
\cdots \lambda _{\tilde p}}=
  {\frac{(-1)^{\frac{N}{2}}N!}{2^{(N+1)}
(2\pi )^{\frac N 2}\left(\frac{N}{2}\right)!}}
\varepsilon_{AA_1A_2\cdots {A_{N-1}A_N}} \\  & \displaystyle \cdot
\frac{\varepsilon ^{\lambda  \lambda _1 \cdots \lambda _{\tilde p} \mu \mu
_1 \cdots \mu _{N}}}{\sqrt{g}} F^{AB}_{\lambda \mu }n^B
F^{{A}_1{A}_2}_{\mu _1\mu _2 } \cdots F^{{A}_{N-1}{A}_N}_{\mu
_{N-1}\mu _{N}} ,
\end{array}
\end{equation}
where $B, A, A_1, A_2, \cdots ,{A_{N-1}, A_N}$ are valued to  $SO(N+1)$ Lie algebra index. Let
\begin{eqnarray}
\Delta ^B = && \varepsilon_{AA_1A_2\cdots {A_{N-1}A_N}}
\frac{\varepsilon ^{\lambda  \lambda _1 \cdots \lambda _{\tilde p} \mu \mu
_1 \cdots \mu _{N}}}{\sqrt{g}} \nonumber  \\ && \cdot
F^{AB}_{\lambda  \mu } F^{{A}_1{A}_2}_{\mu _1\mu _2 } \cdots
F^{{A}_{N-1}{A}_N}_{\mu _{N-1}\mu _{N}} ,
\end{eqnarray}
for fixed $B$, if $A=B$, it's obvious that $\Delta ^B=0$; then one
of $A_1, A_2, \cdots ,{A_{N-1}, A_N}$ must be  valued to $B$  and
\begin{eqnarray}
\Delta ^B = && \varepsilon_{AA_1A_2\cdots {A_{N-1}A_N}}
\frac{\varepsilon ^{\lambda  \lambda _1 \cdots \lambda _{\tilde p} \mu \mu
_1 \cdots \mu _{N}}}{\sqrt{g}}  \nonumber \\ && \cdot
F^{AB}_{\lambda \mu } \cdots F^{{A}_iB}_{\mu _i\mu _{i+1} } \cdots
F^{{A}_{N-1}{A}_N}_{\mu _{N-1}\mu _{N}} ,
\end{eqnarray}
or
\begin{eqnarray}
\Delta ^B = && \varepsilon_{AA_1A_2\cdots {A_{N-1}A_N}}
\frac{\varepsilon ^{\lambda  \lambda _1 \cdots \lambda _{\tilde p} \mu \mu
_1 \cdots \mu _{N}}}{\sqrt{g}} \nonumber \\ && \cdot
F^{AB}_{\lambda  \mu } \cdots F^{B{A}_{i+1}}_{\mu _i\mu _{i+1} }
\cdots F^{{A}_{N-1}{A}_N}_{\mu _{N-1}\mu _{N}} ,
\end{eqnarray}
where $F^{AB}_{\lambda  \mu }$ and $F^{{A}_iB}_{\mu _i\mu _{i+1}
}$, or $ F^{AB}_{\lambda  \mu }$ and $F^{B{A}_{i+1}}_{\mu _i\mu
_{i+1} } $ can be exchanged symmetrically, but the exchange
between $A$ and $A_i$ or $A$ and $A_{i+1}$ is antisymmetric, so we
obtain that $\Delta ^B$ equals to zero also. Therefore, the
$SO(N+1)$ topological tensor current   $\tilde j^{ \lambda \lambda
_1 \cdots \lambda _{\tilde p}} $ is identically conserved
\begin{equation}
 \nabla _{\lambda } \displaystyle {\tilde j}^{\lambda  \lambda _1 \cdots \lambda _{\tilde p}}=0.
\end{equation}
In the case of odd $N$ the dual field tensor $~^{*}F^{\mu \mu _1\cdots \mu _{\tilde p+1}} $ in (\ref{TCCTM}) can be directly expressed in term of the field tensor $F^{AB}_{\mu \nu }$ as expression (\ref{ChernformcompdualN+1}).

As above, the eq.(\ref{TopoTenCurrSON+1}) can be locally written in a simple form
\begin{eqnarray}\label{TopoTenCurrSON+12}
&& {\tilde j}^{\lambda  \lambda _1 \cdots \lambda _{\tilde p}}  = \frac{1}{
2A(S^{N-1})(N-1)!} \varepsilon _{AA_1 A_2 \cdots A_N } \nonumber
\\ && \cdot \frac{\varepsilon ^{\lambda \lambda _1 \cdots \lambda _{\tilde p} \mu \mu
_1 \cdots \mu _{N}}}{\sqrt{g}} \partial _{\mu } n^{A }
\partial _{\mu _1 } n^{A_1 } \partial _{\mu _2 }n ^{A_2 } \cdots
\partial _{\mu _N}n ^{A_{N} }.
\end{eqnarray}

Multipling (\ref{TopoTenCurrSON+1}) and (\ref{TopoTenCurrSON+12})
by ${2A(S^{N-1})}/{NA(S^N)}$ and defining
\begin{equation}\label{}
d= \left\{ \begin{array}{ll}N,&for \quad d \quad even
\vspace{.4cm} \\ N+1,&for \quad d \quad old \end{array} \right.
\end{equation}
where $N$ is even and $D=d+\tilde p+1$ is the dimension of total manifold $X$, then we can obtain a unified topological tensor current on $D$--dimensional smooth manifold $X$
\begin{eqnarray}\label{SO(d)TTC}
\tilde j^{\lambda \lambda _1 \cdots \lambda _{\tilde p}} = &&
 \frac{1}{A(S^{d-1})(d-1)!} \varepsilon_{A_1A_2\cdots
{A_{d-1}A_d}}\nonumber
\\ && \cdot   \frac{\varepsilon ^{ \lambda \lambda _1 \cdots
\lambda _{\tilde p}  \mu _1 \mu _2 \cdots \mu _d}}{\sqrt{g}} \partial _{\mu
_1}n^{A_1}\cdots  \partial _{\mu _d}n^{A_d}. \end{eqnarray}
Here it's easier to prove that above topological currents are locally
conserved
\begin{equation}\label{}
\nabla _{\lambda } {\tilde j}^{\lambda  \lambda _1 \cdots
\lambda _{\tilde p}}=\frac{1}{\sqrt{g}}\partial _{\lambda
}\left(\sqrt{g}{\tilde j}^{\lambda  \lambda _1 \cdots \lambda
_{\tilde p}}\right)=0.
\end{equation}

It's well known that the $\phi $-mapping is a $d$-dimensional smooth vector field on $X$
\begin{equation} \label{phi}
\phi ^A=\phi ^A(x),\quad A=1,2,\cdots ,d,
\end{equation}
and the direction  field of $\vec \phi (x)$ is
\begin{equation} \label{directionfield}
n^A=\frac{\phi ^A}{||\phi ||},\quad \quad ||\phi
||=\sqrt{\phi ^A\phi ^A},
\end{equation}
i.e., $n^A$ is the section of  sphere bundle $S^{d-1}(X)$. Substituting (\ref{directionfield}) into (\ref{SO(d)TTC}) and considering that
\begin{equation}
\partial _\mu n^A=\frac{\partial _\mu \phi ^A}{||\phi ||}+\phi ^A \partial
_\mu \left( \frac 1{||\phi ||}\right) ,
\end{equation}
we have
\begin{equation}   \small
\begin{array}{rl}
\tilde j^{\lambda \lambda _1 \cdots \lambda _{\tilde p}} (x)  = &
\displaystyle\frac{1}{ A(S^{d-1})(d-1)!} \varepsilon _{A_1 \cdots
A_d } \frac{\varepsilon ^{\lambda \lambda _1 \cdots \lambda _{\tilde p} \mu
_1 \cdots \mu _d }}{\sqrt{g}}
\\   & \displaystyle  \cdot \partial _{\mu _1}\left( \frac{\phi ^{A_1 }}{ \left\| \phi
\right\|^d}  \partial _{\mu _2}\phi ^{A_2}\cdots  \partial _{\mu
_d}\phi ^{A_d}\right) \\   = & \displaystyle \frac{1}{
A(S^{d-1})(d-1)!} \varepsilon _{A_1 \cdots A_d } \frac{\varepsilon
^{\lambda \lambda _1 \cdots \lambda _{\tilde p} \mu _1 \cdots \mu _d
}}{\sqrt{g}}
\\   & \displaystyle \cdot  \partial _{\mu _1}\left( \frac{\phi ^{A_1 }}{ \left\| \phi
\right\|^d} \right) \partial _{\mu _2}\phi ^{A_2}\cdots  \partial
_{\mu _d}\phi ^{A_d}.
\end{array}
\end{equation}
Defining the $rank$--$(\tilde p+1)$ Jacobian tensor $J^{\lambda \lambda
_1 \cdots \lambda _{\tilde p}} \left( {{\phi \over x}} \right)$ of $\vec \phi$
as
\begin{equation}
\begin{array}{ll}
& \varepsilon ^{\lambda \lambda _1 \cdots \lambda _{\tilde p} \mu _1 \cdots
\mu _d} \partial _{\mu _1 } \phi ^A\partial _{\mu _2 } \phi ^{A_2
} \cdots \partial _{\mu _d } \phi ^{A_d }  \\ = &  J^{\lambda
\lambda _1 \cdots \lambda _{\tilde p}} \left( {{\phi \over x}}
\right)\varepsilon ^{AA_2 \cdots A_d }, \label{JacobiaofPhi}
\end{array}
\end{equation}
and noticing
\begin{equation}
\varepsilon _{A_1 A_2 \cdots A_d } \varepsilon ^{AA_2 \cdots A_d } =
\delta _{A_1 }^A (d- 1)!,
\end{equation}
it follows that
\begin{equation}
\tilde  j^{\lambda \lambda _1 \cdots \lambda _{\tilde p} } (x) =\frac{1}{
A(S^{d-1})\sqrt{g}} {\partial \over {\partial \phi ^A }}\left(
\frac{\phi ^{A }}{ \left\| \phi \right\|^d} \right) J^{\lambda
\lambda _1 \cdots \lambda _{\tilde p} } \left( {{\phi \over x}} \right).
\label{monopolecurr2}
\end{equation}
Using the Green functions in $\vec \phi $--space
\begin{equation}
{{\phi ^A } \over {\left\| \phi \right\|^d }} = \left\{
\begin{array}{ll}
- {1 \over {(d- 2)}}{\partial \over {\partial \phi ^A }}\left( {{1
\over {\left\| \phi \right\|^{d - 2} }}} \right) & \;\;for \;\; d>2  ,\vspace{.4cm} \\
{\partial \over {\partial \phi ^A }} ln\left\| \phi \right\| &\;\;
for \;\;d=2,
\end{array} \right.
\end{equation}
and
\begin{equation}
\Delta _\phi \left( {{1 \over {\left\| \phi \right\|^{d - 2} }}}
\right) = - \left( {d - 2} \right)A\left( {S^{d - 1} }
\right)\delta \left( {\vec \phi } \right),
\end{equation}
\begin{equation}
\Delta _\phi \left( ln \left\| \phi \right\| \right) = 2\pi \delta
\left( {\vec \phi } \right),
\end{equation}
where $\Delta _{\phi} =\frac{\partial ^2}{\partial \phi ^A\partial
\phi ^A} $ is the $d$-dimensional Laplacian operator in $\vec \phi $
space, we obtain a $\delta $-function like topological tensor
current
\begin{equation} \label{current}
\tilde j^{\lambda \lambda _1 \cdots \lambda
_{\tilde p}}=\frac{1}{\sqrt{g}}\delta (\vec \phi )J^{\lambda \lambda _1
\cdots \lambda _{\tilde p}}(\frac \phi x),
\end{equation}
and find that $ \tilde j^{\lambda \lambda _1 \cdots \lambda
_{\tilde p}}\neq 0$ only when $\vec \phi =0$. So, it is essential to
discuss the solutions of the equations
\begin{equation}
\phi ^A(x)=0,\quad A=1,\cdots ,d.
\end{equation}
This kind of solution plays an crucial role in realization of the
$\tilde p$--brane scenario.

Suppose that the vector field $\vec
\phi (x)$ possesses $\ell$ isolated zeroes, according to the
deduction of Ref.\cite{DuanLiuHopkins1987} and the implicit
function theorem\cite{Goursat04,yang76}, when the zeroes are
regular points of $\phi $--mapping, i.e., the rank of the Jacobian
matrix $[\partial _\mu \phi ^A]$ is $ d$, the solutions of $\vec
\phi (x)=0$ can be parameterized as
\begin{equation}\label{solutdi}
x^\mu =z_i^\mu (u^0,u^1,\cdots ,u^{\tilde p}),\quad i=1,\cdots ,\ell, \quad \mu =1,\cdots ,D,
\end{equation}
where the subscript $i$ represents the $i$--th solution and the
parameters $u^I (I=0,1,2,\cdots ,\tilde{p})$ span a $(\tilde p+1)$--dimensional submanifold  which is called the $i$--th singular submanifold $N_i$ in the total spacetime mainfold $X$. These spatial $\tilde p$--dimension  singular submanifolds $N_i$ are just the
world volumes of the topological  $\tilde p$--branes  in
$M$--theory. The number of solutions $\ell$ takes the role of the
brane number. This is the novel result of the present
work. Therefore by making use of $GBC$ theorem and $\phi
$--mapping topological current theory, we have established a novel
field theory of creating topological $\tilde p$--branes. We must
pointed out that based on the description of $\tilde p$--branes as
topological defects in space-time\cite{DiamantiniPLB1996,duff},
the vector field $\phi ^A(x)$ $(A=1,\cdots ,d)$ can be looked upon
as the order parameter fields of $\tilde p$--branes.

\section{Inner topological structure of $\tilde p$--branes}

From above discussions, we see that the kernel of $\phi
$--mapping plays an crucial role in creating of topological
$\tilde p$--branes. Here we will focus on the zero points of order parameter field $\vec
\phi $ and will search for the inner topological structure of
$\tilde{p}$--branes. It
can be proved that there exists a $d$-dimensional submanifold
$M_i$ in $X$ with local parametric equation
\begin{equation}
\label{trcol}x^\mu =x^\mu (v^1,\cdots ,v^d),\quad \mu =1,\cdots
,D,
\end{equation}
which is transversal to every $N_i$ at the point $p_i$ with metric
\begin{equation}\label{normal}
g_{\mu \nu }\left.B^{\mu }_{I}B^{\nu }_{a}\right|_{p_i} =0, \quad I=0,1,\cdots ,p,\quad a=1,\cdots ,d,
\end{equation}
where
\begin{equation}
\frac{\partial x^\mu }{\partial u^I}=B^{\mu }_{I}  , \quad \frac{\partial x^\nu}{\partial v^a} =B^{\nu }_{a},\quad  \quad  \mu ,\nu =1,2,\cdots ,D,
\end{equation}
are tangent vectors of $N_i$ and $M_i$ respectively. As we have pointed out in Ref.\cite{DuanMengJMP1993}, the unit vector field defined in
(\ref{directionfield}) gives a Gauss map $n:\partial M_i\rightarrow
S^{d-1}$, and the generalized Winding Number can be given by the  map
\begin{equation}\label{}
W_i  = \frac 1{A(S^{d-1})(d-1)!}\int_{\partial M_i}n^{*}(\varepsilon _{A_1\cdots A_d}n^{A_1}dn^{A_2}\wedge \cdots \wedge dn^{A_d}).
\end{equation}
where $n^*$ denotes the pull back of map $n$ and  $\partial M_i$ is the boundary of the neighborhood $M_i$ of
$p_i$ on $X$ with $p_i\notin \partial M_i,$ $M_i\cap
M_j=\emptyset $. It means that, when the point $v^b$ covers $\partial M_i$ once, the unit vector $n$ will cover the region $\partial M_i$ whose area is $W_i$ times of $A^{(d-1)}$, i.e., the unit vector $n$ will cover the unit sphere $A^{(d-1)}$ for $W_i$ times. Using the Stokes' theorem in exterior differential form and  duplicating the derivation of (\ref{current}), we obtain
\begin{equation} \label{wind}
W_i=\int_{M_i}\delta (\vec \phi (v))J(\frac \phi v)d^dv
\end{equation}
where $J(\frac \phi v)$ is the usual Jacobian determinant of $\vec
\phi $ with respect to $v$
\begin{equation}
\varepsilon ^{A_1\cdots A_d}J(\frac \phi v)=\varepsilon ^{\mu
_1\cdots \mu _d}\partial _{\mu _1}n^{A_1}\partial _{\mu
_2}n^{A_2}\cdots \partial _{\mu _d}n^{A_d}.
\end{equation}
According to the $\delta $-function theory\cite{Schouten} and the
$\phi $--mapping theory, we know that $\delta (\vec \phi)$ can be expanded as
\begin{equation} \label{ff}
\delta (\vec \phi )=\sum_{i=1}^\ell c_i \delta (N_i)
\end{equation}
where the coefficients $c_i$ must be positive, i.e., $c_i=|c_i|$. $\delta (N_i)$ is the $\delta $--function in $X$ on a submanifold $N_i$,\cite{Schouten,Gelfand}
\begin{equation}
\delta (N_i)=\int_{N_i}\delta ^D(x-z_i(u))\sqrt{g_u}d^{(\tilde p+1)}u,\quad
\quad i=1,\cdots ,\ell ,
\end{equation}
where $g_u=det(g_{IJ})$. Substituting (\ref{ff}) into (\ref{wind}), and calculating the integral, we get the expression of $c_i$,
\begin{equation}\label{}
c_i=\frac{\beta _i}{\left|J \left(\frac{\phi }{v}\right)\right|_{p_i}}=\frac{\beta _i\eta _i}{\left. J \left(\frac{\phi }{v}\right)\right|_{p_i}},
\end{equation}
where the positive integer $\beta _i=|W_i|$ is called
the Hopf index of $\phi $--mapping on $M_i$, and $\eta
_i=sgn(J(\frac \phi v))|_{p_i}=\pm 1$ is the Brouwer
degree\cite{DuanMengJMP1993,Hopf}. So we find the relations between the
Hopf index $\beta _i,$ the Brouwer degree $ \eta _i$, and the
winding number $W_i$
\begin{equation}
W_i=\beta _i\eta _i.
\end{equation}
Therefore, the general topological current of the $ \tilde
p$--branes can be expressed directly as
\begin{equation}
\begin{array}{rl} \label{recurdi}
\tilde j^{\lambda \lambda _1 \cdots \lambda _{\tilde p}} = &
\frac{1}{\sqrt{g}}J^{\lambda \lambda _1 \cdots \lambda
_{\tilde p}}(\frac{\phi }{x})
\\  & \cdot \sum_{i=1}^{\ell }\beta _i\eta _i\int_{N_i}
\delta ^D(x-z_i(u))\sqrt{g_u}d^{(\tilde p+1)}u.
\end{array}
\end{equation}
From the above equation, we conclude that the $(\tilde p+1)$-dimensional singular submanifolds
$N_i\;(i=1,2,\cdots ,\ell) $ are world volumns  of $\tilde p$--branes, and the inner topological structure of $\tilde p$--branes  current  is labelled by the total expansion of $\delta (\vec \phi)$  which includes the topological information $\beta _i \eta _i$. In detail, $\beta _i$ characterizes the absolute value of the topological charge of every $\tilde p$--brane and $\eta _i=+1$ corresponds to $\tilde p$--brane while $\eta _i=-1$ to anti $\tilde p$--brane.  Taking the parametre $u^0$ and $u^I \;\; (I=1,2,\cdots , \tilde p)$   as the timelike evolution parameter and spacelike parameters respectively,  the topological current of $\tilde p$--branes  just represents $\tilde p$--dimensional topological defects with topological charges
$\beta _i \eta _i$ moving in the $D$--dimensional total manifold
$X$.

If we define a Lagrangian of $\tilde p$--brane as
\begin{equation}\label{GenNieLag}
\label{nestring}L=\sqrt{\frac 1{(\tilde p+1)!}g_{\mu _0\nu _0}g_{\mu _1\nu _1}\cdots
g_{\mu _{\tilde p}\nu _{\tilde p}} \tilde j^{\mu _0 \mu _1\cdots \mu _{\tilde p}}
\tilde j^{\nu _0\nu _1\cdots \nu _{\tilde p}}},
\end{equation}
which is just the generalization of Nielsen's Lagrangian of
string\cite{niel} and includes the total information of arbitrary dimensional $\tilde p$--branes in $X$. It obvious that the Euler equations corresponding to the Lagrangian  (\ref{GenNieLag}) will give the dynamics of the $\tilde{p}$-branes. From the above deductions, we can prove that
\begin{equation}
L=\frac 1{\sqrt{g}}\delta (\vec \phi (x)).
\end{equation}
Then, the action takes the form
\begin{eqnarray}\label{aa}
S && =\int_XL\sqrt{g}d^Dx =\int_X\delta (\vec \phi (x))d^Dx
\nonumber \\ && =\sum_{i=1}^l\beta _i\eta
_i\int_{N_i}\sqrt{g_u}d^{(\tilde p+1)}u,
\end{eqnarray}
i.e.
\begin{equation}
S=\sum_{i=1}^l\eta _iS_i,
\end{equation}
where $S_i=\beta _i\int_{N_i}\sqrt{g_u}d^{(\tilde p+1)}u.$ This is just
the straightforward generalized Nambu action for the
string world-sheet action \cite{nambu}. Here this action for multi $\tilde p$--branes is obtained directly by $\phi $-mapping
theory, and it is easy to see that this action is just Nambu action for multistrings when $D-d=2$.\cite{DuanLiuHopkins1987}.


\begin{thebibliography}{99} \addcontentsline{toc}{section}{Bibliography}
\small \parskip-2mm \addcontentsline{toc}{section}{Bibliography}
\bibitem{b1}  J. H. Schwarz, hep-th/9607201.
\bibitem{b2}  E. Witten, Nucl. Phys. {\bf B 443}, 85(1995).
\bibitem{b3}  P. K. Townsend, Phys. Lett. {\bf B 350}, 184(1995).
\bibitem{b4}  C. M. Hull, Nucl. Phys. {\bf B 468}, 113(1996).
\bibitem{StelleTownsend1987} K.S. Stelle and P.K. Townsend, "Are 2-branes better than 1?" in Proc. CAP Summer Institute, Edmonton, Alberta, July 1987, KEK library accession number 8801076.
\bibitem{SchwarzSeibergRMP1999} J. H. Schwarz and N. Seiberg, Rev. Mod. Phys. \textbf{71}, S112(1999).
\bibitem{DiamantiniPLB1996}M. C. Diamantini, Phys. Lett. \textbf{B388}, 273(1996), arXiv:hep-th/9607090.
\bibitem{a1}  J. Scherk an J. H. Schwarz, Phys. Lett. {\bf B52}, 347(1974); J.H. Schwarz, Phys. Rep. {\bf 89}, 223(1982).
\bibitem{a2}  Y. Nambu, Phys. Rep. {\bf 23}, 250(1976).
\bibitem{a3}  R. I. Nepomechie, Phys. Rev. {\bf D31}, 1921(1985).
\bibitem{strom}  A. Strominger, Nucl. Phys. {\bf B 343}, 167(1990).
\bibitem{duff}  M. J. Duff, R.R. Khuri and J. X. Lu, Phys. Rep. {\bf 256}, 213(1995).
\bibitem{hull}  C. M. Hull, Nucl. Phys. {\bf B509}, 216(1998).
\bibitem{DvaliPRD2000} G. R. Dvali, I. I. Kogan, and M. A. Shifman, Phys. Rev., \textbf{D62}, 106001(2000).
\bibitem{CembranosPRD2002} J. A. R. Cembranos, A. Dobado, A. L. Maroto, Phys.Rev.\textbf{D65}, 026005(2002); L. Perivolaropoulos, hep-ph/0307269.
\bibitem{DuanSlac1984} Y. S. Duan, SLAC-PUB-3301/84.
\bibitem{DuanRenYang2003} Y. S. Duan, J. R. Ren  J. and Yang, Chinese Phys. Lett. {\bf 20},  2133(2003).
\bibitem{DuanLiuJHEP2004} Y. S. Duan and X. Liu, J. HIGH ENERGY PHYS. \textbf{A041}, 0403(2004).
\bibitem{DuanMengJMP1993}  Y. S. Duan, X. H. Meng , J. Math. Phys., {\bf 34}, 1149(1993).
\bibitem{DuanLiYangNPB1998}  Y. S. Duan, S. Li, G. H. Yang, Nucl. Phys. {\bf B 514}, 705(1998).
\bibitem{DuanFuJMP1998}  Y. S. Duan, L. B. Fu, J. Math. Phys., {\bf 39}, 4343 (1998).
\bibitem{DuanLiuFuPRD2003}  Duan Y. S., Liu X. and  Fu L. B., Phys. Rev. \textbf{D67}, 085022(2003).
\bibitem{DUANZHangLiPRB1998}Y. S. Duan, H. Zhang, and S. Li, Phys. Rev. \textbf{B58}, 125(1998).
\bibitem{DUANLiuZHangJP2002} Y. S. Duan, X. Liu and P. M. Zhang, J. Phys.: Condens. Matter \textbf{14}, 7941(2002).
\bibitem{DuanZhangFuPRE1999} Y. S. Duan, H. Zhang and L. B. Fu, Phys. Rev. \textbf{E59}, 528(1999).
\bibitem{DuanZhangPRE1999} Y. S. Duan, and H. Zhang, Phys. Rev.
{\bf E 60}, 2568(1999).
\bibitem{DuanFuZhangPRD2000}  L. B. Fu, Y. S. Duan, H. Zhang,Phys. Rev.  {\bf D61}, 045004(2000).
\bibitem{DuanFuJiaJMP2000} Y. S. Duan; L. B. Fu; G. Jia, J. Math. Phys., \textbf{41}, 4379(2000).
\bibitem{DuanLiuHopkins1987} Y. S. Duan and J. C. Liu, \textit{ Proceedings of the 11th Johns Hopkins Workshop on Current Problem in Particle Theory}, \textbf{11}(1987)183, Lanzhou, China, edited by Y. S. Duan, G. Domokos, and S. Kovesi-Domokos, World Scientific, Singapore, (1988).
\bibitem{DuanZhangMPLA2001} Y. S. Duan and P. M. Zhang, Mod. Phys. Lett. \textbf{A16}, 2483(2001).
\bibitem{JiangDuanJMP2000} Y. Jiang, Y. S. Duan, J. Math. Phys. \textbf{41},  6463(2000).
\bibitem{YangJiangDuanCPL2001}G. H. Yang, Y. Jiang, Y. S. Duan, Chin. Phys. Lett. \textbf{18}, 631(2001).
\bibitem{NepomechiePRD1985} R. I. Nepomechie, Phys. Rev. \textbf{D31}, 1921(1985).
\bibitem{Cher1} S. S. Chern, Ann. Math. {\bf 45}, 747(1944); {\bf{46}}, 674(1945); S.S. Chern, {\it Differentiable
Manifolds.} Lecture Notes Univ. of Chicago, (1959).
\bibitem{Dowk}  J. S. Dowker, and J. P. Schofield, J. Math. Phys. {\bf31}, 808(1990).
\bibitem{Alva}  Alvarez-Gaum\'e, L.  Commun. Math. Phys. {\bf 90}, 161(1983).
\bibitem{Weissteink-Form} Eric W. Weisstein et al. "Differential k-Form." From MathWorld--A Wolfram Web Resource. http://mathworld.wolfram.com/Differentialk-Form.html
\bibitem{spivak1975} Spivak M., Differential Geometry, Publish or Perish, Boston. (1975).
\bibitem{Goursat04}  \'{E}. Goursat, {\it A Course in Mathematical Analysis,} translated by Earle Raymond Hedrick, (Ginn \& Company, Boston, 1905) Vol. I.
\bibitem{yang76}  L. V. Toralballa, {\it Theory of Functions}, (Charies E. Merrill Books, Inc., Columbus, Ohio, (1963).
\bibitem{Schouten}  J. A. Schouten, {\it Tensor Analysis for Physicists}, Oxford, Clarendon Press, (1951).
\bibitem{Gelfand} I.M. Gelfand and G.E. Shilov, \textit{Generalized Function} National Press of Mathematics Literature, Moscow, (1958)
\bibitem{Hopf} H. Hopf, Math. Ann. \textbf{96}, 209(1929).
\bibitem{niel}  H. B. Nielsen and P. Olesen, Nucl. Phys. {\bf B57},
367(1973).
\bibitem{nambu}  Y. Nambu, {\it Lectures at the Copenhagen Symposium}, 1970.
\end{thebibliography}
\end{document}